\documentclass[11pt]{article}

\usepackage{epsf}
\usepackage{color}
\usepackage{amsmath,amssymb,color,graphics,amscd,amsfonts}

\begin{document}

\begin{flushright} 
February, 2013 \\
OIQP-13-03
\end{flushright} 

\vspace{0.1cm}

\begin{Large}
\vspace{1cm}
\begin{center}
{\bf Non-planar operator mixing by Brauer representations} \\ 
\end{center}
\end{Large}

\vspace{0.2cm}

\begin{center}
{\large Yusuke Kimura}\footnote{
Okayama Institute for Quantum Physics (OIQP), Kyoyama 1-9-1, Kita-ku, Okayama, 700-0015, Japan;  
londonmileend@gmail.com}
\\ 

\end{center}

\vspace{0.8cm}

\begin{abstract}
\noindent 
We study the action of the dilatation operator 
on the basis of local operators 
constructed 
from elements of 
the walled Brauer algebra, with non-planar corrections 
fully taken into account. 
We will see that the operator mixing 
can be neatly expressed in terms of 
the irreducible representations of the algebra. 
In particular  
we focus on a role of the integer that determines the number of boxes 
in the representations.

\end{abstract}

\vspace{0.5cm}


\quad

\section{Introduction}

In 4D ${\cal N}=4$ Super Yang-Mills theory, 
conformal symmetry is a strong symmetry which we can use to restrict 
the form of correlation functions. As for two-point functions 
one has the form
\begin{eqnarray}
\langle O_i (x)O_j(y)\rangle=N_i \delta_{ij}(x-y)^{-\Delta_i}, 
\end{eqnarray}
where $\Delta_i$ is the scaling dimension of the local operator. 
However,  
to obtain the scaling dimension in concrete cases  
we have to look for a good basis 
from the naive basis so that two-point functions 
have the diagonal form. 
(See  
\cite{0109064,0205321,0206020,0307081} for example). 
In other words 
the good operators are eigenstates of the dilatation operator 
with an eigenvalue $\Delta_i$ 
\cite{0212208,0303060}. 

Resolving the operator mixing problem needs a hard job in general, 
but 
as far as the planar limit is concerned,   
we have had a big progress 
since the discovery of an integrable structure underlying 
the theory
\cite{0212208,1012.3982}. 
In the planar theory only single-trace operators come in the game, 
which are mapped to spin states. 
Considering only single-traces 
is justified by restricting 
our attention to operators 
whose classical dimension is much smaller than 
the size of the gauge group $N$. 
On the other hand, 
we need a new perspective if our focus is on large operators 
with a classical dimension comparable to $N$ or more, 
where 
huge combinatoric factors arising
from summing up non-planar diagrams spoil the planar approximation. 
Consider a single-trace operator $tr(X^{J})$. 
It is given 
a good interpretation
as a dual string object if $J \ll N$. In contrast, 
an issue arises if we consider a single-trace operator 
with $J\sim O(N)$. 
For example  
$tr(X^{N+1})$ 
can be expressed 
as a linear combination of smaller operators. 
This implies that  
it should not correspond 
to an independent vertex operator in string theory.  
The equation to express $tr(X^{N+1})$ in terms of smaller operators 
can be obtained from 
the character of $SU(N)$ as 
$\chi_R (X)=0$, where $R=[1^{N+1}]$ is the anti-symmetric representation 
with $N+1$ boxes. 
This fact may encourage us to make use of Young diagrams for such large operators, 
and in fact 
Young diagrams can be used to count the number of 
large operators \cite{0111222}. 
The constraint that 
Young diagrams should have no more than $N$ rows plays a role 
\cite{0107119}. 
In the non-planar regime
the operator corresponding to an independent vertex operator of 
string theory is a linear combination of multi-trace operators. 
Such large operators are considered to be dual to giant gravitons or geometries 
\cite{0003075,0403110,0409174}. 

The finite $N$ issue for multi-matrix 
can also be understood in terms of Young diagrams. 
Some complete sets of local gauge invariant operators 
were constructed 
which are built from 
some kinds of fields 
\cite{0709.2158,0711.0176,0801.2061,0805.3025,0807.3696}.  
They are labelled by a set of Young diagrams, and 
their notable property is that 
the operators have diagonal two-point functions 
at free theory, which are exact with respect to $N$. 
Such diagonal bases can be a good starting point to study 
the operator mixing problem for large operators.  
There have been some studies along this line to find that 
the operator mixing is neatly handled in terms of Young diagrams
 \cite{0701066,0710.5372,0801.2094,1002.2099,1002.2424,1004.1108,1010.1683,
1012.3884,1101.5404,1108.2761,1111.1058,1204.2153,
1206.0813,1207.6948,1212.6624}. 
See \cite{0303060,0404066,0710.4166,1012.3997} 
for former approaches to non-planar corrections. 

In this paper, we study the action of 
the dilatation operator on the Brauer basis introduced in \cite{0709.2158}. 
The basis 
has some labels, and one of the labels is an 
irreducible representation of the walled Brauer algebra $B_N(m,n)$
\cite{Stembridge,Koike1989,BCHLLS}, which is given by 
a bi-partition,
\begin{eqnarray}
\gamma=(\gamma_+,\gamma_-).
\end{eqnarray}
Here\footnote{
$R\vdash m$ is a shorthand notation to express that 
the Young diagram $R$ contains $m$ boxes.
}  
\begin{eqnarray}
\gamma_+ \vdash (m-k),\quad \gamma_-\vdash (n-k) , 
\end{eqnarray}
where $k$ is an integer obeying $0 \le k \le min(m,n) $. 
We also denote it by $\gamma=(\gamma_+,\gamma_-,k)$ 
to emphasise what $k$ it has 
because 
we will pay attention to a role of the integer $k$.\footnote{
On a personal note the reason we focus on $k$ is as follows.
For a class of 1/4 BPS operators constructed in \cite{1002.2424}, 
a dual meaning of the integer $k$ was conjectured 
in the context of the correspondence with 
the 1/4 BPS geometries in \cite{1109.2585}. 
In our previous paper \cite{1206.4844} 
we have studied correlation functions of Brauer operators 
and observed that correlation functions may be classified 
in terms of the integer $k$. 
}
We will show how the two operators  
get mixed 
under the action of the dilatation operator. 
Particularly we will indicate the mixing condition for $k$. 
The mixing matrix we will obtain is expressed in terms of 
representation date 
and is very constrained.  
In other words, the mixing is very restricted on the representations. 
It would be 
a universal property for the diagonal bases labelled by 
a set of Young diagrams.

The paper is organised as follows. 
In section \ref{sec:one-loop}, 
we will study the one-loop dilatation operator on the Brauer basis. 
We follow the idea in \cite{1012.3884} used to evaluate 
the dilatation operator 
on the restricted Schur basis.
In section \ref{mixing_one-loop}, we 
will write down 
the mixing condition 
in terms of the Brauer representations. 
In section \ref{sec:two-loop}, 
\ref{sec:mixing_two-loop}, we provide with the same study for 
two-loop. 
In section \ref{general_action}, 
the action of more general differential operators 
are studied. 
Section \ref{summary} is given for summary and discussions. 
Appendix \ref{basic} is for basic things of Brauer bases.

\section{One-loop dilatation operator on Brauer basis}
\label{sec:one-loop}

The operator is 
\begin{eqnarray}
O^{\gamma}_{A,ij}(X,Y)
=tr_{m,n}(Q^{\gamma}_{A,ij}X^{\otimes m}\otimes Y^{T \otimes n}).
\label{TheBasis}
\end{eqnarray}
See appendix \ref{basic} for a detailed explanation. 

The one-loop dilatation operator is given by 
$\hat{H}_2=-tr([X,Y][\partial_X,\partial_Y])$ \cite{0303060}, which acts on 
the Brauer basis 
as 
\begin{eqnarray}
\hat{H}_2 O^{\gamma}_{A,ij}
&=&
-
mn
tr_{m,n}\left([Q^{\gamma}_{A,ij},C_{mn}]
 X^{\otimes m-1}\otimes [X,Y]
\otimes Y^{T \otimes n-1}\otimes 1\right)
\nonumber \\
&=&
-
mn
\sum_{r,kl}
\frac{1}{d_r}\chi^{\gamma}_{r,kl}\left([Q^{\gamma}_{A,ij},C_{mn}]\right)
tr_{m,n}\left(Q^{\gamma}_{r,kl}
 X^{\otimes m-1} \otimes [X,Y]
\otimes Y^{T \otimes n-1} \otimes 1\right),
\label{action_dilatation_Brauer}
\end{eqnarray}
where 
$r$ is an irreducible representation of the group algebra of 
$S_{m-1}\times S_1\times 
S_{n-1}\times S_1$. 
To obtain the second line we have used 
the inverse formula
(\ref{inverse_formula}). 
The contraction is denoted by $C_{mn}$, which acts on the $m$-th slot of 
$X^{\otimes m}$ and the $n$-th slot of $Y^{T\otimes n}$.

We now recall the result of \cite{1002.2424}, where it was shown 
that the following operator 
\begin{eqnarray}
O^{\gamma}(X,Y)=tr_{m,n}(P^{\gamma}X^{\otimes m}\otimes Y^{T \otimes n}),
\label{1/4bps_opes}
\end{eqnarray}
where $P^{\gamma}=\sum_{A,i}Q^{\gamma}_{A,ii}$ is 
the projection operator associated with 
an irreducible representation $\gamma$, 
is in the kernel of the dilatation operator. 
The result is very manifest in the form of (\ref{action_dilatation_Brauer}) 
because of $[P^{\gamma},C_{mn}]=0$.

\quad

Following 
the strategy 
in \cite{1012.3884}, we will 
expand (\ref{action_dilatation_Brauer}) in terms of the basis (\ref{TheBasis}).  
The readers who are not interested in the derivation can jump over to 
(\ref{mixing_finalform_one-loop}).
Expressing the elements in $B_N(m,n)$ 
in terms of $B_N(m,n-1)$ as in 
(\ref{B(m,n)toB(m,n-1)}), 
we have\footnote{
We have  the following equations 
for $c\in B_N(m,n-1)$;
\begin{eqnarray}
&&
tr_{m,n}(c
 X^{\otimes m-1} \otimes [X,Y]
\otimes Y^{T \otimes n-1} \otimes 1)
=
N
tr_{m,n-1}(c
 X^{\otimes m-1} \otimes [X,Y]
\otimes Y^{T \otimes n-1}),
\nonumber \\
&&
tr_{m,n}((\bar{i},\bar{n})c
X^{\otimes m-1} \otimes [X,Y]
\otimes Y^{T \otimes n-1} \otimes 1)
=
tr_{m,n-1}(c
 X^{\otimes m-1} \otimes [X,Y]
\otimes Y^{T \otimes n-1}),
\nonumber \\
&&
tr_{m,n}(cC_{kn}
X^{\otimes m-1} \otimes [X,Y]
\otimes Y^{T \otimes n-1} \otimes 1)
=
tr_{m,n-1}(c
 X^{\otimes m-1} \otimes [X,Y]
\otimes Y^{T \otimes n-1}),
\end{eqnarray}
where 
$i=1,\cdots,n-1$ and 
$k=1,\cdots,m$. 
}

\begin{eqnarray}
&&
tr_{m,n}(Q^{\gamma}_{r,kl}
X^{\otimes m-1} \otimes [X,Y]
\otimes Y^{T \otimes n-1} \otimes 1)
\nonumber \\
&=&
t^{\gamma}
\sum_{b\in B_N(m,n)}
\chi^{\gamma}_{r,lk}(b^{\ast})
tr_{m,n}(b
X^{\otimes m-1} \otimes [X,Y]
\otimes Y^{T \otimes n-1} \otimes 1)
\nonumber \\
&=&
t^{\gamma}
\sum_{c\in B_N(m,n-1)}
\chi^{\gamma}_{r,lk}(O_c^{\ast})
tr_{m,n-1}(c
 X^{\otimes m-1} \otimes [X,Y]
\otimes Y^{T \otimes  n-1}), 
\label{rewriting}
\end{eqnarray}
where 
\begin{eqnarray}
O_c^{\ast}
:=Nc^{\ast}+\sum_{i=1}^{n-1}((\bar{i},\bar{n})c)^{\ast}
+\sum_{k=1}^{m}(cC_{kn})^{\ast}. 
\end{eqnarray}
The first equality in 
(\ref{rewriting}) comes from the definition of $Q^{\gamma}_{r,kl}$ 
(see (\ref{def_Q})). 
Note that $c^{\ast}$ is not an element in $B_N(m,n-1)$ but 
an element in $B_N(m,n)$.
Thanks to the formula (\ref{the_formula}) we find in 
appendix \ref{reduction_formula}, 
(\ref{rewriting}) can be rewritten as 
\begin{eqnarray}
t^{\gamma}
\sum_{c\in B_N(m,n-1)}
\chi^{\gamma}_{r,lk}(c^{\tilde{\ast}})
tr_{m,n-1}(c
 X^{\otimes m-1} \otimes [X,Y]
\otimes Y^{T \otimes  n-1}).
\end{eqnarray}
Here $c^{\tilde{\ast}}$ is an element in $B_N(m,n-1)$.
Using the following formula
\begin{eqnarray}
tr_{m,n-1}(c
 X^{\otimes m-1} \otimes [X,Y]
\otimes Y^{T \otimes n-1})=
tr_{m,n}([C_{mn},c]
 X^{\otimes m}
\otimes Y^{T\otimes  n})
\label{aformula_commutator}
\end{eqnarray}
for $c \in B_{N}(m,n-1)$ and the inverse formula  
(\ref{inverse_formula_2matrix}), we can expand the action of 
the one-loop dilatation operator in terms of the basis 
as 
\begin{eqnarray}
\hat{H}_2 O^{\gamma}_{A,ij}=
\sum_{\gamma^{\prime},A^{\prime},pq}
M_{(\gamma,A,ij)}^{(\gamma^{\prime},A^{\prime},pq)}
O_{A^{\prime},pq}^{\gamma^{\prime}},
\end{eqnarray}
where the mixing matrix is 
\begin{eqnarray}
M_{(\gamma,A,ij)}^{(\gamma^{\prime},A^{\prime},pq)}
=
-
mn
\sum_{r,kl}
\frac{1}{d_r}\chi^{\gamma}_{r,kl}([Q^{\gamma}_{A,ij},C_{mn}])
t^{\gamma}
\sum_{c\in B_N(m,n-1)}
\chi^{\gamma}_{r,lk}(c^{\tilde{\ast}})
\frac{1}{d_{A^{\prime}}}
\chi_{A^{\prime},pq}^{\gamma^{\prime}}
([C_{mn},c]).
\label{eq1}
\end{eqnarray}
Making use of the formula (\ref{character_combine})
\footnote{
Note that $[Q^{\gamma}_{A,ij},C_{mn}]$ commutes with any elements 
in $S_{m-1}\times S_1 \times S_{n-1}\times S_1 $.
}, 
the two characters can be combined to give 
\begin{eqnarray}
M_{(\gamma,A,ij)}^{(\gamma^{\prime},A^{\prime},pq)}
=
-
mn
t^{\gamma}
\sum_{c\in B_N(m,n-1)}
\chi^{\gamma}([Q^{\gamma}_{A,ij},C_{mn}]c^{\tilde{\ast}})
\frac{1}{d_{A^{\prime}}}
\chi_{A^{\prime},pq}^{\gamma^{\prime}}
([C_{mn},c]).
\label{eq2}
\end{eqnarray}
We next perform the sum over $c$
using the orthogonality relation of the algebra.
In order to use the orthogonality
\footnote{
The orthogonality relation of the Brauer algebra is given by
\begin{eqnarray}
\sum_{c\in B_N(m,n-1)} \Gamma^{(\gamma_1)}(c)_{\alpha\beta}
\Gamma^{(\gamma_1^{\prime})}(c^{\tilde{\ast}})_{\gamma\delta}
=\frac{1}{t^{\gamma_1}} \delta_{\alpha\delta}
\delta_{\beta\gamma} \delta^{\gamma_1 \gamma_1^{\prime}},
\end{eqnarray}
where $\gamma_1$, $\gamma_1^{\prime}$
are irreducible representations of $B_N(m,n-1)$.
}
 for $B_N(m,n-1)$ 
we have to consider 
the restriction from $B_N(m,n)$
to a subalgebra $B_N(m,n-1)$. 
Let 
$\gamma_1$ be one of the irreducible representations 
that appear 
in $\gamma$ upon restricting $B_N(m,n)$ to $B_N(m,n-1)$, 
and $\gamma_1^{\prime}$ likewise.
Then the sum can be evaluated in the following way
using intertwiners 
\begin{eqnarray}
&&
\sum_{c\in B_N(m,n-1)}
\chi^{\gamma}([Q^{\gamma}_{A,ij},C_{mn}]c^{\tilde{\ast}})
\chi_{A^{\prime},pq}^{\gamma^{\prime}}
([C_{mn},c])
\nonumber \\
&=&
\sum_{c\in B_N(m,n-1)}
\Gamma^{\gamma}([Q^{\gamma}_{A,ij},C_{mn}])_{ba}\Gamma^{\gamma}(c^{\tilde{\ast}})_{ab}
\Gamma^{\gamma^{\prime}}
([Q_{A^{\prime},qp}^{\gamma^{\prime}},C_{mn}])_{dc}
\Gamma^{\gamma^{\prime}}(c)_{cd}
\nonumber \\
&=&
\sum_{\gamma_1,\gamma_1^{\prime}}
\frac{1}{t^{\gamma_1}}
tr\left(
\Gamma^{\gamma}([Q^{\gamma}_{A,ij},C_{mn}])
I_{\gamma_1\gamma_{1}^{\prime}}
\Gamma^{\gamma^{\prime}}
\left([Q_{A^{\prime},qp}^{\gamma^{\prime}},C_{mn}]\right)
I_{\gamma_{1}^{\prime}\gamma_1}
\right).
\end{eqnarray}
$I_{\gamma_1\gamma_{1}^{\prime}}$ is an intertwiner that 
maps from $\gamma_1$ to $\gamma_1^{\prime}$, 
and it is non-zero if 
 $\gamma_1$ and $\gamma_1^{\prime}$ are the same shape. 
See appendix B in \cite{1012.3884} or appendix D in \cite{1108.2761} 
for more details on the intertwiners. 

Then the final form of our mixing matrix is 
\begin{eqnarray}
M_{(\gamma,A,ij)}^{(\gamma^{\prime},A^{\prime},pq)}
=
-
mn
\frac{1}{d_{A^{\prime}}}
\sum_{\gamma_1,\gamma_1^{\prime}}
\frac{t^{\gamma}}{t^{\gamma_1}}
\chi^{\gamma}\left(
[Q^{\gamma}_{A,ij},C_{mn}]
I_{\gamma_1\gamma_{1}^{\prime}}
[Q_{A^{\prime},qp}^{\gamma^{\prime}},C_{mn}]
I_{\gamma_{1}^{\prime}\gamma_1}
\right).
\label{mixing_finalform_one-loop}
\end{eqnarray}
We have obtained a very similar form to 
the case of restricted Schur basis. 
The $N$-dependence is exact.

We find that 
the BPS operators (\ref{1/4bps_opes})
do not appear 
even in the image of the dilatation operator:
\begin{eqnarray}
\langle O^{\gamma^{\prime\prime}}
\hat{H}_2 O_{A,ij}^{\gamma}\rangle
&=&\sum_{p}\sum_{A^{\prime\prime}}
M^{(\gamma^{\prime\prime},A^{\prime\prime},pp)}_{(\gamma,A,ij)}
m!n!d_{A^{\prime \prime}}t^{\gamma^{\prime\prime}}
\nonumber \\
&\sim &
\chi^{\gamma}(\cdots [P^{\gamma^{\prime\prime}},C_{mn}]\cdots)
=0, 
\end{eqnarray}
where we have used the two-point function (\ref{free_two-point}). 
Therefore they are completely separated from the others.

\quad

The first line of (\ref{action_dilatation_Brauer})
may be written as 
\begin{eqnarray}
\hat{H_2} O^{\gamma}_{A,ij}
&=&
-
mn tr_{m,n}(
Q^{\gamma}_{A,ij}C_{mn}
X^{\otimes m-1}\otimes XY
\otimes Y^{T}{}^{\otimes n-1} \otimes 1 )
\nonumber \\
&&
+
mn
tr_{m,n}(Q^{\gamma}_{A,ij}C_{mn} X^{\otimes m}\otimes Y^{T}{}^{\otimes n})
\nonumber \\
&&
+
mn
tr_{m,n}(C_{mn} Q^{\gamma}_{A,ij}X^{\otimes m}\otimes Y^{T}{}^{\otimes n})
\nonumber \\
&&
-
mn
tr_{m,n}(
C_{mn}
Q^{\gamma}_{A,ij}
X^{\otimes m-1}\otimes YX
\otimes Y^{T}{}^{\otimes n-1} \otimes 1 ).
\end{eqnarray}
This leads to another form of 
the mixing matrix:
\begin{eqnarray}
M_{(\gamma,A,ij)}^{(\gamma^{\prime},A^{\prime},pq)}
&=&
-
mn
\frac{1}{d_{A^{\prime}}}
\sum_{\gamma_1,\gamma_1^{\prime}}
\frac{t^{\gamma}}{t^{\gamma_1}}
\chi^{\gamma}\left(
Q^{\gamma}_{A,ij}C_{mn}
I_{\gamma_1\gamma_{1}^{\prime}}
Q_{A^{\prime},qp}^{\gamma^{\prime}}C_{mn}
I_{\gamma_{1}^{\prime}\gamma_1}
\right)
\nonumber \\
&&
-mn
\frac{1}{d_{A^{\prime}}}
\sum_{\gamma_1,\gamma_1^{\prime}}
\frac{t^{\gamma}}{t^{\gamma_1}}
\chi^{\gamma}\left(
C_{mn}Q^{\gamma}_{A,ij}
I_{\gamma_1\gamma_{1}^{\prime}}
C_{mn}Q_{A^{\prime},qp}^{\gamma^{\prime}}
I_{\gamma_{1}^{\prime}\gamma_1}
\right)
\nonumber \\
&&
+\delta_{\gamma}^{\gamma^{\prime}}\delta_{A}^{A^{\prime}}
\delta_{i}^{p}mn
\frac{1}{d_A}\chi^{\gamma}_{A,jq}(C_{mn})
+\delta_{\gamma}^{\gamma^{\prime}}\delta_{A}^{A^{\prime}}
\delta_{j}^{q}mn
\frac{1}{d_A}\chi^{\gamma}_{A,pi}(C_{mn}).
\label{mixing-2ndform}
\end{eqnarray}


\section{One-loop mixing}
\label{mixing_one-loop}

The one-loop mixing matrix we have obtained 
can be shown to be in fact  
almost diagonal. 
The necessary condition to 
have a non-zero mixing is that  
$\gamma_1$ and $\gamma_1^{\prime}$ are the same bi-partition. 
In this section 
we will translate this condition into a condition between 
$\gamma$ and $\gamma^{\prime}$. 

The restriction of $B_N(m,n)$ to $B_N(m,n-1)$ was studied in 
\cite{1206.4844}. 
An irreducible representation $\gamma_1$ in $B_N(m,n-1)$ 
can appear in 
an irreducible representation $\gamma$ in $B_N(m,n)$ 
if the following quantity is non-zero \cite{Koike1989} 
\begin{eqnarray}
M_{\gamma\rightarrow \gamma_1}=\sum_{\delta,\zeta}
g(\delta,\gamma_+;\gamma_{1+})
g(\delta,\zeta;[1])
g(\zeta,\gamma_{1-};\gamma_{-}),
\end{eqnarray}
where $g(\alpha,\beta;\gamma)$ is the Littlewood-Richardson 
coefficient. 
We now introduce a new notation 
\begin{eqnarray}
R=S^{(+1)}
\end{eqnarray}
to denote that $R$ is a Young diagram obtained from 
$S$ by adding a box. We also denote it by 
$S=R^{(-1)}$. 
For instance $R^{(-1)}=R^{\prime (-1)}$ means that 
$R$ and $R^{\prime}$ are related each other by moving a single box.
In terms of this new notation, 
there are two cases for $M_{\gamma\rightarrow \gamma_1}$ to be non-zero:
\begin{eqnarray}
\gamma_+=\gamma_{1+},\quad \gamma_{-}=\gamma_{1-}^{(+1)}, 
\end{eqnarray}
which is the case $(\delta,\zeta)=(\emptyset,[1])$, 
and 
\begin{eqnarray}
\gamma_+^{(+1)}=\gamma_{1+},\quad \gamma_{-}=\gamma_{1-}, 
\end{eqnarray}
which is the case
$(\delta,\zeta)=([1],\emptyset)$. 
Introducing $\Delta=k-k_1$, 
the first case is $\Delta=0$ while the second is 
$\Delta=1$.

We find that 
to get a non-zero mixing element 
we need one of the following conditions between   
$\gamma$ and $\gamma^{\prime}$:
\begin{itemize}
\item $(\Delta,\Delta^{\prime})=(0,0)$ 
\begin{eqnarray}
\gamma_+=\gamma_+^{\prime}, 
\quad \gamma_{-}^{(-1)}=\gamma_{-}^{\prime}{}^{(-1)}
\qquad [k=k^{\prime}],
\end{eqnarray}
\item $(\Delta,\Delta^{\prime})=(1,1)$
\begin{eqnarray}
\gamma_+^{(+1)}=\gamma_+^{\prime}{}^{(+1)},
\quad \gamma_{-}=\gamma_{-}^{\prime}
\qquad [k=k^{\prime}],
\end{eqnarray}
\item $(\Delta,\Delta^{\prime})=(0,1)$
\begin{eqnarray}
\gamma_+=\gamma_+^{\prime}{}^{(+1)}, 
\quad \gamma_{-}^{(-1)}=\gamma_{-}^{\prime}
\qquad [k=k^{\prime}-1],
\end{eqnarray}

\item $(\Delta,\Delta^{\prime})=(1,0)$
\begin{eqnarray}
\gamma_+^{(+1)}=\gamma_+^{\prime}, 
\quad \gamma_{-}=\gamma_{-}^{\prime}{}^{(-1)} 
\qquad [k=k^{\prime}+1]. 
\end{eqnarray}

\end{itemize}

The one-loop dilatation operator
induces an interaction 
between two sets of Young diagrams that are related by moving a box. 
In this sense, we say 
that only ``nearest-neighbour" Young diagrams interact each 
other at one-loop. 
This seems to be an interesting extension of the planar case, 
where the one-loop dilatation 
operator was a nearest-neighbour interaction on spin states. 
This kind of nearest-neighbour interactions on Young diagrams 
have been already seen on the other bases 
\cite{0801.2094,1012.3884}.  
We suspect that in our case 
the integer $k$ is a convenient index to 
know if two operators are nearest-neighbours.


\section{Two-loop dilatation operator on Brauer basis}
\label{sec:two-loop}
In this section, we 
will evaluate the action of the two-loop dilatation operator 
on the Brauer basis. 
The two-loop dilatation operator on the restricted Schur basis 
was evaluated in \cite{1206.0813}. 

The two-loop dilatation operator was given in \cite{0303060} as
\begin{eqnarray}
\hat{H}_4
&=&
-tr(:[[Y,X],\partial_X][[\partial_Y,\partial_X],X]:)
-tr(:[[Y,X],\partial_Y][[\partial_Y,\partial_X],Y]:)
\nonumber \\
&&
-tr(:[[Y,X],T_a][[\partial_Y,\partial_X],T_a]:). 
\label{two-loop-dilatation}
\end{eqnarray}
The last term is identical to 
the one-loop dilatation operator. 
Because the second term is obtained by exchanging $X$ and $Y$ 
in the first term, we will study only the first term.

After some straightforward calculation, which we will show in 
appendix \ref{two-loop_detail}, we obtain 
\begin{eqnarray}
&& 
\frac{1}{m(m-1)n}
tr(:[[Y,X],\partial_X][[\partial_Y,\partial_X],X]:)O^{\gamma}_{A,ij}
\nonumber \\
&=&
tr_{m,n}\left([C_{mn}(m-1,m),Q^{\gamma}_{A,ij}] 
X^{\otimes m-2}\otimes 
X[Y,X]\otimes 1\otimes 
Y^{T}{}^{\otimes n-1}\otimes 1\right)
\nonumber \\
&&
-
tr_{m,n}\left([(m-1,m)C_{mn},Q^{\gamma}_{A,ij}]  
X^{\otimes m-1} \otimes [Y,X]\otimes 
Y^{T}{}^{\otimes n-1}\otimes 1 \right)
\nonumber \\
&&
-
tr_{m,n}\left([C_{mn}(m-1,m),Q^{\gamma}_{A,ij}] 
 X^{\otimes m-1}\otimes [Y,X]\otimes 
Y^{T}{}^{\otimes n-1} \otimes 1\right)
\nonumber \\
&&
+
tr_{m,n}\left([(m-1,m)C_{mn},Q^{\gamma}_{A,ij}] 
X^{\otimes m-2}\otimes 
[Y,X]X\otimes 1\otimes 
Y^{T}{}^{\otimes n-1}\otimes 1
\right).
\label{first_terms_D4}
\end{eqnarray}
The second and the third term above
are similar to the one-loop result.
We further rewrite the first term and the forth term,  
which we need the reduction formula twice.

Note that in the form it is pretty manifest that 
the operators 
(\ref{1/4bps_opes}) are BPS at two-loop. 
It is interesting to observe that $Q^{\gamma}_{A,ij}$ always appears 
in the form $[Q^{\gamma}_{A,ij},b]$ with an element $b$ in the algebra.

We now concretely show how the first term 
in (\ref{first_terms_D4})
can be further rewritten. 
Using the inverse formula 
it can be expanded in terms of 
\begin{eqnarray}
tr_{m,n}\left(Q^{\gamma}_{s,kl}  X^{\otimes m-2}\otimes X[Y,X]\otimes
1\otimes  
Y^{T}{}^{\otimes n-1}\otimes 1\right), 
\end{eqnarray}
where $s$ is an irreducible representation of the group algebra of
$S_{m-2}\times S_1\times S_1\times S_{n-1}\times S_1$. 
After using the formula given 
in appendix \ref{reduction_formula} twice, the above operator 
can be rewritten as 
\begin{eqnarray}
&&
t^{\gamma}\sum_{b\in B_N(m,n)}
\chi^{\gamma}_{s,lk}(b^{\ast}) 
tr_{m,n}\left(b X^{\otimes m-2}\otimes X[Y,X]\otimes
1\otimes  
Y^{T}{}^{\otimes n-1}\otimes 1\right)
\nonumber \\
&=&
t^{\gamma}\sum_{c\in B_N(m-1,n-1)}
\chi^{\gamma}_{s,lk}(c^{\tilde{\ast}}) 
tr_{m-1,n-1}\left(c X^{\otimes m-2}\otimes X[Y,X] \otimes  
Y^{T}{}^{\otimes n-1}\right). 
\end{eqnarray}
Here the dual element 
$c^{\tilde{\ast}}$ is defined in $B_N(m-1,n-1)$. 
We can use the following equations for $c\in B_{N}(m-1,n-1)$: 
\begin{eqnarray}
tr_{m-1,n-1}(c X^{\otimes m-2}\otimes XYX \otimes  
Y^{T}{}^{\otimes n-1})
&=&tr_{m,n}((m-1,m) C_{mn}c X^{\otimes m}\otimes 
Y^{T}{}^{\otimes n}),
\nonumber \\
tr_{m-1,n-1}(c X^{\otimes m-2}\otimes XXY \otimes  
Y^{T}{}^{\otimes n-1})
&=&tr_{m,n}( C_{mn}(m-1,m)c X^{\otimes m}\otimes 
Y^{T}{}^{\otimes n}),
\nonumber \\
tr_{m-1,n-1}(c X^{\otimes m-2}\otimes YXX \otimes  
Y^{T}{}^{\otimes n-1})
&=&tr_{m,n}(c(m-1,m)C_{mn} X^{\otimes m}\otimes 
Y^{T}{}^{\otimes n}). 
\end{eqnarray}
Note that if we use $[c,C_{mn}]=0$ for $c\in B_{N}(m-1,n-1)$ 
and $(m-1,m) X^{\otimes m}=X^{\otimes m}(m-1,m)$, 
there are several equivalent expressions like 
\begin{eqnarray}
tr_{m,n}((m-1,m) C_{mn}c 
X^{\otimes m}\otimes Y^{T}{}^{\otimes n}) 
=
tr_{m,n}(c C_{mn} (m-1,m) 
X^{\otimes m}\otimes Y^{T}{}^{\otimes n}). 
\end{eqnarray}
Then the first term in (\ref{first_terms_D4}) is
\begin{eqnarray}
&&
tr_{m,n}\left( [C_{mn}(m-1,m),Q^{\gamma}_{A,ij}] 
X^{\otimes m-2}\otimes 
1\otimes X[Y,X]\otimes 
Y^{T}{}^{\otimes n-1} \otimes 1 \right)
\nonumber \\
&=&
\sum_{s,kl}
\frac{1}{d_s}\chi_{s,kl}^{\gamma}\left([C_{mn}(m-1,m),Q^{\gamma}_{A,ij} ]
\right)
tr_{m,n}\left(Q_{s,kl}^{\gamma} 
X^{\otimes m-2}\otimes 
1\otimes X[Y,X]\otimes 
Y^{T}{}^{\otimes n-1} \otimes 1 
\right)
\nonumber \\
&=&
\sum_{s,kl}
\frac{1}{d_s}\chi_{s,kl}^{\gamma}\left(
[C_{m,n}(m-1,m),Q^{\gamma}_{A,ij} ]\right)
\nonumber \\
&& \qquad \times 
t^{\gamma}\sum_{c\in B(m-1,n-1)} 
\chi_{s,lk}^{\gamma}(c^{\tilde{\ast}})
tr_{m-1,n-1}\left(c X[Y,X]
\otimes X^{\otimes m-2}\otimes 
Y^{T}{}^{\otimes n-1}\right)
\nonumber \\
&=&
\sum_{s,kl}
\frac{1}{d_s}\chi_{s,kl}^{\gamma}\left(
[C_{mn}(m-1,m),Q^{\gamma}_{A,ij} ]\right)
\nonumber \\
&& \qquad \times 
t^{\gamma}\sum_{c\in B(m-1,n-1)} 
\chi_{s,lk}^{\gamma}(c^{\tilde{\ast}})
tr_{m,n}\left([c,C_{mn}(m-1,m)] 
 X^{\otimes m}\otimes 
Y^{T}{}^{\otimes n}\right)
\nonumber \\
&=&
\sum_{s,kl}
\frac{1}{d_s}\chi_{s,kl}^{\gamma}
\left([C_{mn}(m-1,m),Q^{\gamma}_{A,ij} ]\right)
t^{\gamma}\sum_{c\in B(m-1,n-1)} 
\chi_{s,lk}^{\gamma}(c^{\tilde{\ast}})
\nonumber \\
&&\qquad 
\times 
\sum_{\gamma^{\prime},A^{\prime},pq}
\frac{1}{d_{A^{\prime}}}
\chi^{\gamma^{\prime}}_{A^{\prime},pq}\left([c,C_{mn}(m-1,m)] \right)
tr_{m,n}(Q^{\gamma^{\prime}}_{A^{\prime},pq}
\otimes X^{\otimes m}\otimes 
Y^{T}{}^{\otimes n}) .
\end{eqnarray}
Making use of 
(\ref{character_combine}) and 
performing the sum of $c \in B_N(m-1,n-1)$, 
we have 
\begin{eqnarray}
\hat{H}_4 O^{\gamma}_{A,ij}=
M_{(\gamma,A,ij)}^{(\gamma^{\prime},A^{\prime},pq)}
O^{\gamma^{\prime}}_{A^{\prime},pq},
\end{eqnarray}
where the contribution from the first term in (\ref{first_terms_D4})
to $M_{(\gamma,A,ij)}^{(\gamma^{\prime},A^{\prime},pq)}$
is 
\begin{eqnarray}
m(m-1)n
\sum_{\gamma_2,\gamma_2^{\prime}} 
\frac{t^{\gamma}}{t^{\gamma_2}}
\frac{1}{d_{A^{\prime}}}
\chi^{\gamma}\left(
[C_{mn}(m-1,m),Q^{\gamma}_{A,ij} ]
I_{\gamma_2 \gamma_2^{\prime}} 
[C_{mn}(m-1,m),Q^{\gamma^{\prime}}_{A^{\prime},qp}]
I_{\gamma_2^{\prime} \gamma_2} \right),
\end{eqnarray}
where $\gamma_2$ and $\gamma_2^{\prime}$ 
are irreducible representations of $B_N(m-1,n-1)$. 
Likewise 
the forth term in (\ref{first_terms_D4}) gives
\begin{eqnarray}
m(m-1)n
\sum_{\gamma_2,\gamma_2^{\prime}} 
\frac{t^{\gamma}}{t^{\gamma_2}}
\frac{1}{d_{A^{\prime}}}
\chi^{\gamma}\left(
[(m-1,m)C_{mn},Q^{\gamma}_{A,ij} ]
I_{\gamma_2 \gamma_2^{\prime}} 
[(m-1,m)C_{mn},Q^{\gamma^{\prime}}_{A^{\prime},qp}]
I_{\gamma_2^{\prime} \gamma_2} \right).
\end{eqnarray}
From these expressions 
we find that 
the BPS operators (\ref{1/4bps_opes})
do not appear in the image under the two-loop dilatation operator.


\section{Two-loop mixing}
\label{sec:mixing_two-loop}
The necessary condition for 
the two-loop mixing matrix to be non-zero 
is that 
$\gamma_2$ and $\gamma_2^{\prime}$ are the same shape. 
The mixing structure is determined by 
the restriction $B_N(m,n)\rightarrow 
B_N(m-1,n-1)\times B_N(1,1)$. 
The number of times $\gamma_2$ appears in an irreducible representation $\gamma$ 
is counted by 
\begin{eqnarray}
&& M_{\gamma_2,\gamma_3}^{\gamma}
=
\sum_{\rho,\zeta,\theta,\kappa}
\left(
\sum_{\delta}
g(\delta,\rho;\gamma_{2+})
g(\delta,\zeta;\gamma_{3-})
\right)
\left(
\sum_{\epsilon}
g(\epsilon,\theta;\gamma_{2-})
g(\epsilon,\kappa;\gamma_{3+})
\right) \nonumber \\
&& \hspace{4cm}\times 
g(\rho,\kappa;\gamma_{+})
g(\zeta,\theta;\gamma_{-}),
\label{branching_number}
\end{eqnarray}
where we have denoted an irreducible representation of 
$B_N(1,1)$ by $\gamma_3$. 
There are two irreducible representations 
in $B_N(1,1)$;
$\gamma_3=([1],[1],k_3=0)$ or $\gamma_3=(\emptyset,\emptyset,k_3=1)$. 
For the former case, 
we have 
\begin{eqnarray}
M_{\gamma_2,\gamma_3 (k_3=0)}^{\gamma}
=
\sum_{\rho,\zeta,\theta,\kappa}
\left(
\sum_{\delta}
g(\delta,\rho;\gamma_{2+})
g(\delta,\zeta;[1])
\right)
\left(
\sum_{\epsilon}
g(\epsilon,\theta;\gamma_{2-})
g(\epsilon,\kappa;[1])
\right)
g(\rho,\kappa;\gamma_{+})
g(\zeta,\theta;\gamma_{-}),
\nonumber 
\end{eqnarray}
while for the latter case we have 
\begin{eqnarray}
M_{\gamma_2,\gamma_3 (k_3=1)}^{\gamma}
&=&
\sum_{\rho,\zeta,\theta,\kappa}
\left(
\sum_{\delta}
g(\delta,\rho;\gamma_{2+})
g(\delta,\zeta;\emptyset)
\right)
\left(
\sum_{\epsilon}
g(\epsilon,\theta;\gamma_{2-})
g(\epsilon,\kappa;\emptyset)
\right)
g(\rho,\kappa;\gamma_{+})
g(\zeta,\theta;\gamma_{-})
\nonumber \\
&=&
g(\gamma_{2+},\emptyset;\gamma_{+})
g(\emptyset,\gamma_{2-};\gamma_{-}). 
\nonumber 
\end{eqnarray}
In these cases the 
$M_{\gamma_2,\gamma_3}^{\gamma}$ takes 0 or 1. 

We find that there are five cases 
the $M_{\gamma_2,\gamma_3}^{\gamma}$ takes the non-zero value.
Introducing $\Delta=k-k_2-k_3$, 
we list the five cases below: 
\begin{itemize}
\item 
$\gamma_{2+}=\gamma_+^{(-1)}$, $\gamma_{2-}=\gamma_-^{(-1)}$ \qquad 
[$k_3=0$, $k=k_2$, $\Delta=0$], 

\item 
$\gamma_{2+}=\gamma_+$, $\gamma_{2-}=\gamma_-$
\qquad [$k_3=1$, $k=k_2+1$, $\Delta=0$],

\item 
$\gamma_{2+}=\gamma_+$, $\gamma_{2-}^{(-1)}=\gamma_-^{(-1)}$ 
\qquad [$k_3=0$, $k=k_2+1$, $\Delta=1$],

\item 
$\gamma_{2+}^{(-1)}=\gamma_+^{(-1)}$, $\gamma_{2-}=\gamma_-$ 
\qquad [$k_3=0$, $k=k_2+1$, $\Delta=1$],

\item 
$\gamma_{2+}=\gamma_+^{(+1)}$, $\gamma_{2-}=\gamma_-^{(+1)}$ 
\qquad [$k_3=0$, $k=k_2+2$, $\Delta=2$].

\end{itemize}
They correspond to 
$(\delta,\zeta,\epsilon,\kappa)=(0,[1],0,[1]),(0,0,0,0),
(0,[1],[1],0),([1],0,0,[1]),([1],0,[1],0)$
respectively. 

Because there are five cases for the restriction 
from $\gamma^{\prime}$ to $\gamma_2^{\prime}$ as well, 
we have twenty-five cases 
$\gamma$ and $\gamma^{\prime}$ interact each other 
in which $k$ and $k^{\prime}$ are related by  
$k^{\prime}=k,k\pm1,k\pm2$. 
We will show some of them: 
\begin{itemize}
\item $(\Delta,\Delta^{\prime})=(0,0)$, $(k_3,k_3^{\prime})=(0,0)$
\begin{eqnarray}
\gamma_+^{(-1)}=\gamma_+^{\prime}{}^{ (-1)}, 
\quad \gamma_{-}^{(-1)}=\gamma_{-}^{\prime}{}^{(-1)} \qquad [k=k^{\prime}],
\end{eqnarray}
\item $(\Delta,\Delta^{\prime})=(0,2)$, $(k_3,k_3^{\prime})=(1,0)$
\begin{eqnarray}
\gamma_+=\gamma_+^{\prime}{}^{ (+1)}, 
\quad \gamma_{-}=\gamma_{-}^{\prime}{}^{(+1)} \qquad [k=k^{\prime}-1],
\end{eqnarray}
\item $(\Delta,\Delta^{\prime})=(2,0)$, $(k_3,k_3^{\prime})=(0,0)$
\begin{eqnarray}
\gamma_+^{(+1)}=\gamma_+^{\prime}{}^{ (-1)}, 
\quad \gamma_{-}^{(+1)}=\gamma_{-}^{\prime}{}^{(-1)} \qquad [k=k^{\prime}+2],
\end{eqnarray}
\item $(\Delta,\Delta^{\prime})=(0,0)$, $(k_3,k_3^{\prime})=(1,1)$
\begin{eqnarray}
\gamma_+=\gamma_+^{\prime}, 
\quad \gamma_{-}=\gamma_{-}^{\prime} \qquad [k=k^{\prime}],
\end{eqnarray}
\item $(\Delta,\Delta^{\prime})=(1,1)$, $(k_3,k_3^{\prime})=(0,0)$
\begin{eqnarray}
&&
\gamma_+=\gamma_+^{\prime}, 
\quad \gamma_{-}^{(-2)}=\gamma_{-}^{\prime}{}^{(-2)} \qquad [k=k^{\prime}],
\nonumber \\
&&
\gamma_+^{(-2)}=\gamma_+^{\prime}{}^{(-2)}, 
\quad \gamma_{-}=\gamma_{-}^{\prime} \qquad [k=k^{\prime}],
\nonumber \\
&&
\gamma_+^{(-1)}=\gamma_+^{\prime}{}^{(-1)}, 
\quad \gamma_{-}^{(-1)}=\gamma_{-}^{\prime}{}^{(-1)} \qquad [k=k^{\prime}],
\end{eqnarray}
where the last one appears twice.
\end{itemize}

\section{More general differential operators}
\label{general_action}
In this section we will 
study the action of 
more general differential operators on the basis 
with 
focusing on the role of $k$. 
Let us first consider the differential operator 
$tr(W\partial_X \partial_X\partial_Y\partial_Y)$ as an example, 
assuming that $W$ is made from two copies of $X$ 
and two copies of $Y$. 
The action on the basis can be evaluated as 
\begin{eqnarray}
&&
tr(W\partial_X \partial_X\partial_Y\partial_Y) O^{\gamma}_{A,ij}(X,Y)
\nonumber \\
&=&
m(m-1)n(n-1)
tr_{m,n}\left(Q^{\gamma}_{A,ij} C_{2,1}(12)(\bar{1}\bar{2})
W\otimes 1\otimes X^{\otimes m-2}\otimes 1\otimes 
1\otimes Y^{T}{}^{\otimes n-2} \right)
\nonumber \\
&=&
m(m-1)n(n-1)
\sum_{r,kl}\chi^{\gamma}_{r,kl}
\left(Q^{\gamma}_{A,ij} C_{2,1}(12)(\bar{1}\bar{2})\right)
\nonumber \\
&& \quad \times 
tr_{m,n}\left(Q^{\gamma}_{r,kl}
W\otimes 1\otimes X^{\otimes m-2}\otimes 1\otimes 
1\otimes Y^{T}{}^{\otimes n-2} \right) ,
\end{eqnarray}
where $r$ is an irreducible representation of the group algebra of
$S_{m-2}\times S_1\times S_1\times S_{n-2}\times S_1\times S_1$. 

The next step is to 
use the reduction rule three times, followed by  
taking the sum of $c\in B_N(m-1,n-2)$ appearing in  
$Q^{\gamma}_{r,kl}$. 
Finally, we will have
$\chi^{\gamma}(\cdots I_{\gamma_2\gamma_2^{\prime}} \cdots 
I_{\gamma_2^{\prime}\gamma_2})$,  
where $\gamma_2$ and $\gamma_2^{\prime}$ 
are irreducible representations of $B_N(m-1,n-2)$. 
This leads to considering 
the restriction 
$B_N(m,n)\rightarrow B_N(m-1,n-2)\times B_N(1,2)$, 
which determines the mixing structure. 
Denoting an irreducible representation of $B_N(1,2)$ by $\gamma_3$, 
the necessary condition for the non-zero mixing is that 
the $M_{\gamma_2,\gamma_3}^{\gamma}$, 
which is given by the same form as  
(\ref{branching_number}), 
is non-zero.
Denoting the number of boxes in a partition $\alpha$ by 
$n(\alpha)$, we find that 
$n(\delta)$ and $n(\zeta)$ take $0,1,2$, and 
$n(\epsilon)$ and $n(\kappa)$ take $0,1$. 
From the consistency, we have the relation
\begin{eqnarray}
n(\gamma_+)=n(\kappa)+n(\gamma_{2+})-n(\delta), 
\end{eqnarray}
which implies 
\begin{eqnarray}
k-k_2=0,1,2,3. 
\end{eqnarray}
Therefore the necessary condition for non-zero mixing 
between two representations $\gamma$ 
and $\gamma^{\prime}$ is 
$k^{\prime}=k,k\pm 1,k\pm 2,k\pm 3$. 
Schematically writing, we have 
\begin{eqnarray}
tr(W\partial_X \partial_X\partial_Y\partial_Y)O^{\gamma(k)}\sim \sum_{k^{\prime}=k-3}^{k+3}
a_{k^{\prime}}O^{\gamma^{\prime}(k^{\prime})}.
\end{eqnarray}

If we consider the action of a more general differential operator 
with $p$ $\partial_X$'s and $q$ $\partial_Y$'s, 
we have to consult the restriction 
$B_N(m,n)\rightarrow B_N(m-p+1,n-q)\times B_N(p-1,q)$. 
By a similar argument to the above case, 
we find that 
\begin{eqnarray}
k-k_2=0,1,2,\cdots,p+q-1, 
\end{eqnarray}
which implies 
\begin{eqnarray}
k^{\prime}=k,k\pm 1,\cdots,
k\pm (p+q-1).
\end{eqnarray}


\section{Summary and discussions}
\label{summary}
In this paper we have expressed the following mixing matrix 
in terms of Brauer representation data 
with keeping the $N$-dependence exact,  
\begin{eqnarray}
\langle O^{\gamma^{\prime}}_{A^{\prime},i^{\prime}j^{\prime}}
{}^{\dagger}
\hat{H}_{2l}
O^{\gamma}_{A,ij}
\rangle .
\end{eqnarray}
We have found that the mixing on the representations is highly restrictive. 
Such restricted mixing would be a universal property of 
the bases for which 
the free two-point function is diagonal.  
 
In particular 
we have focused on the integer $k$ that determines the number of 
boxes in the irreducible representation $\gamma$ of the Brauer algebra. 
To get a non-zero mixing result, we need $k-k^{\prime}=0,\pm 1$ at one-loop
and $k-k^{\prime}=0,\pm 1,\pm 2$ at two-loop. 
This mixing is reminiscent of \cite{1004.1108}.  
It would be interesting to consider if 
the dilatation operator 
can be viewed as a lattice approximation to a second derivative on $k$. 
We also have the mixing for the Young diagrams
($\gamma_+$, $\gamma_-$).
The diagrams 
($\gamma_+$, $\gamma_-$) encode the information of $k$ 
but if we consider 
a three-dimensional space
spanned by $\gamma_+$, $\gamma_-$ and $k$,  
we might get a good perspective to understand the mixing. 
In order to gain more understanding of a role of the integer $k$, 
it would also be nice to study the dilatation operator 
on the Brauer bases beyond su(2) sector \cite{1206.4844,0910.2170}.

We have explicitly confirmed 
that the construction of some 1/4 BPS operators 
in \cite{1002.2424} 
is valid at two-loop. We also saw that 
they do not appear in the image of the dilatation operator. 

The next direction we should proceed to is 
to diagonalise the mixing matrix. 
Recently there have been an interesting progress in diagonalising 
the mixing matrix on 
the restricted Schur operators
\cite{1012.3884,1101.5404,1108.2761,1206.0813,1212.6624}. 
They have shown that 
the action of the dilatation operator reduces to systems of 
harmonic oscillators when the corners of the Young diagram are well-separated 
(displaced corner approximation). 
The idea underlying the approximation is to exploit 
a good set of conserved charges 
preserved by the dilatation operator. 
In the series of works, 
the charges encode open string configurations on giant gravitons 
\cite{1111.1058,1204.2153}. 
(Another approach can be seen in \cite{1301.3519}.)
Finding good conserved charges is a nice pathway towards 
diagonalising the mixing matrix.  
Because such conserved charges correspond to parameters 
of the dual physics, it would be helpful to conduct an analysis 
in the string/gravity side. 
In \cite{1109.2585} we have studied a correspondence between 
1/4 BPS operators and 1/4 BPS geometries, where 
it was shown that the geometries are characterised by 
an integer that has the same upper bound as the integer $k$ 
(: recall $k\le min(m,n)$).  
This fact might be a clue 
to solve the operator mixing dual to 
geometries.

\qquad

{\bf Acknowledgements}: 
I would like to thank Robert de Mello Koch and Sanjaye Ramgoolam
for helpful discussions. 
I also would like to thank 
Durham university. Discussions during  
the workshop 
``Symmetry and Geometry of Branes 
in String/M Theory'' were useful. 

\qquad 

\appendix 
\renewcommand{\theequation}
{\Alph{section}.\arabic{equation}}

\section{Basic facts}
\label{basic}
\setcounter{equation}{0}
In this appendix, we summarise 
basic things of Brauer operators, which were 
proposed in \cite{0709.2158} and 
further studied in \cite{0807.3696,1002.2424,1206.4844,0911.4408}. 
The operator we consider can be expressed by 
\begin{eqnarray}
O^{\gamma}_{A,ij}(X,Y)=
tr_{m,n}\left(Q^{\gamma}_{A,ij}X^{\otimes m}\otimes Y^{T \otimes n}
\right),
\end{eqnarray}
where $\gamma=(\gamma_+,\gamma_-)$ is an irreducible representation of 
the walled Brauer algebra $B_N(m,n)$, 
and a pair of Young diagrams $A=(\alpha,\beta)$ 
is an irreducible representation of
$S_m\times S_n$. 
The group algebra of $S_m\times S_n$ is a subalgebra of 
$B_N(m,n)$, and a representation $\gamma$ in general contains 
some representations of the group algebra of $S_m\times S_n$. 
When a particular representation of the group algebra of 
$S_m\times S_n$ appears 
more than once in the restriction, 
the labels $i,j$ specify which copy of the representation 
we consider. 
The multiplicity of the representation $A$ in 
a representation $\gamma$ is given by 
\begin{eqnarray}
M^{\gamma}_A=\sum_{\delta\vdash k}
g(\delta,\gamma_+ ;\alpha)
g(\delta,\gamma_- ;\beta), 
\end{eqnarray}
where 
$g(\delta,\gamma_+ ;\alpha)$ is the Littlewood-Richardson coefficient. 
The labels on the operator are summarised as 
\begin{eqnarray}
&&
\gamma=(\gamma_+,\gamma_-), \quad 
\gamma_+\vdash (m-k), \quad 
\gamma_-\vdash (n-k) 
\nonumber 
\\
&&
A=(\alpha,\beta), \quad 
\alpha\vdash m, \quad 
\beta\vdash n 
\nonumber 
\\
&& i,j =1,\cdots, M^{\gamma}_A ,
\end{eqnarray}
where $k$ is an integer in the range $0\le k \le min(m,n)$. 
The representations should satisfy 
\begin{eqnarray}
c_1(\gamma_+)+c_1(\gamma_-)\le N, \quad 
c_1(\alpha)\le N , \quad 
c_1(\beta) \le N, 
\end{eqnarray}
where $c_1(R)$ denotes the length of the first column of 
the Young diagram $R$. 

The two-point function is diagonal under the free field 
computation 
\begin{eqnarray}
\langle 
O^{\gamma^{\prime}}_{A^{\prime},i^{\prime}j^{\prime}}{}
^{\dagger}
O^{\gamma}_{A,ij}
\rangle
=
m!n!
\delta_{\gamma\gamma^{\prime}}
\delta_{AA^{\prime}}
\delta_{ii^{\prime}}
\delta_{jj^{\prime}} 
d_A t^{\gamma}, 
\label{free_two-point}
\end{eqnarray} 
where the space-time dependence is omitted. 
$t^{\gamma}$ is the dimension of $\gamma$ considered 
as a representation of $U(N)$, 
and $d_A$ is the dimension of $A$ as a representation of $S_m\times S_n$. 

\quad 

If 
we regard the matrices $X$ and $Y$ as linear operators on 
a vector space $V$, the tensor product 
$X^{\otimes m}\otimes Y^{T \otimes n}$ can be regarded 
as a linear operator acting on the vector space 
$W=V^{\otimes m}\otimes \bar{V}^{\otimes n}$.
The Brauer operator is constructed by acting with 
$Q^{\gamma}_{A,ij}$ on the tensor product 
$X^{\otimes m}\otimes Y^{T \otimes n}$ and taking a trace on 
$W$. 
The trace is written by $tr_{m,n}$. 
The $Q^{\gamma}_{A,ij}$ can be expressed as a linear combination of 
elements in the Brauer algebra, which 
can be written explicitly as 
\begin{eqnarray}
 Q^{\gamma}_{A,ij}=t^{\gamma}\sum_{b\in B_N(m,n)}
\chi^{\gamma}_{A,ji}(b^{\ast})b.
\label{def_Q} 
\end{eqnarray} 
The coefficient of the linear combination is 
the restricted character of the Brauer algebra: 
$\chi^{\gamma}_{A,ji}(b)=\chi^{\gamma}(Q^{\gamma}_{A,ij}b)$. 
Another ingredient appearing in 
$Q^{\gamma}_{A,ij}$ is 
$b^{\ast}$, which we call the dual element, is a specific linear combination of 
elements in the algebra. 
In \cite{0709.2158} the following formula 
was given for the dual element:
\begin{eqnarray}
\Sigma( b_i^{\ast})=\frac{1}{N^{m+n}}\Omega_{m+n}^{-1}
(\Sigma(b_i))^{-1},
\label{dual_element}
\end{eqnarray}
using a map $\Sigma:B_N(m,n)\rightarrow S_{m+n}$, 
and $\Omega_{m+n}$ is the Omega factor 
(see \cite{0709.2158} for the definition). 
The index $i$ runs 
over a complete set of 
the elements in 
the algebra. 

One important property of $Q^{\gamma}_{A,ij}$ is 
\begin{eqnarray}
\sigma Q^{\gamma}_{A,ij} \sigma^{-1}=Q^{\gamma}_{A,ij}, \qquad 
\sigma \in S_m\times S_n. 
\end{eqnarray}

The projection operators of the Brauer algebra 
associated with an irreducible representation $\gamma$ are given by  
$P^{\gamma}=\sum_{A,i}Q^{\gamma}_{A,ii}$. 
Note that they are 
in the centre of the algebra. 
If we replace $X$ and $Y$ with a unitary matrix $U$ and 
the conjugate $U^{\dagger}$, 
$O^{\gamma}(U,U^{\dagger})=
tr_{m,n}\left(P^{\gamma}U^{\otimes m}\otimes U^{\ast \otimes n}
\right)$ is the character of $U(N)$. 
So roughly speaking, our operators are obtained by decomposing 
the characters into some pieces. 

We have 
the inverse formula 
\begin{eqnarray} 
tr_{m,n}(bX^{\otimes m}\otimes Y^{T \otimes n}) 
=
\sum_{\gamma,A,ij}
\frac{1}{d_A}
\chi^{\gamma}_{A,ij}(b)tr_{m,n}(Q^{\gamma}_{A,ij}
X^{\otimes m}\otimes Y^{T \otimes n}).
\label{inverse_formula_2matrix}
\end{eqnarray} 
Noticing that 
any multi-trace operator can be written in the form 
of the left-hand side with a certain element $b$ of the algebra, 
this formula enables us to express any 
multi-trace operator in terms of 
the basis. 

For an element $b_0$ satisfying $\sigma b_0 =b_0 \sigma$ for 
any $\sigma \in S_m\times S_n$, we have 
(see appendix B in \cite{0807.3696})
\begin{eqnarray} 
b_0 Q^{\gamma}_{A,ij}=\frac{1}{d_A}\sum_{k}\chi^{\gamma}_{A,ki}(b_0)
Q^{\gamma}_{A,kj}. 
\end{eqnarray} 
This is consistent with the inverse formula. 
Making use of this formula, 
we can derive the following 
\begin{eqnarray} 
\sum_{A,kl}
\frac{1}{d_A}\chi^{\gamma}_{A,kl}(b_0)
\chi^{\gamma}_{A,lk}(b)
&=&
\sum_{A,kl}
\frac{1}{d_A}\chi^{\gamma}_{A,kl}(b_0)
\chi^{\gamma}(Q^{\gamma}_{A,kl} b)
\nonumber \\
&=&
\sum_{A,l}
\chi^{\gamma}(b_0 Q^{\gamma}_{A,ll} b)
\nonumber \\
&=&
\chi^{\gamma}(b_0 b). 
\end{eqnarray} 

\quad 

The above things are for the su(2) sector, 
but it is also possible to accommodate 
more than two fields \cite{1206.4844}. For example, 
the basis 
for $m-s$ $X$'s, $s$ $W_1$'s, $n-t$ $Y$'s and $t$ $W_2$'s 
is given by the same way just by replacing 
$A$ with $r$ that is an irreducible representation of 
$S_{m-s}\times S_s \times S_{n-t}\times S_t$. 
The inverse formula for this case is   
\begin{eqnarray} 
&&
tr_{m,n}(bX^{\otimes m-s}\otimes W_1^{\otimes s}\otimes Y^{T \otimes n- t}
\otimes W_2^{T\otimes t}) \nonumber \\
&=&
\sum_{\gamma,r,ij}
\frac{1}{d_r}
\chi^{\gamma}_{r,ij}(b)tr_{m,n}(Q^{\gamma}_{r,ij}
X^{\otimes m-s}\otimes W_1^{\otimes s}\otimes Y^{T \otimes n- t}
\otimes W_2^{T\otimes t}).
\label{inverse_formula}
\end{eqnarray} 
We also have 
\begin{eqnarray} 
\sum_{r,kl}
\frac{1}{d_r}\chi^{\gamma}_{r,kl}(b_0)
\chi^{\gamma}_{r,lk}(b)
=
\chi^{\gamma}(b_0 b), 
\label{character_combine}
\end{eqnarray} 
where $b_0$ is an element that commutes 
with any elements in $S_{m-s}\times S_s \times S_{n-t}\times S_t$. 


\section{A reduction formula}
\label{reduction_formula}
\setcounter{equation}{0}

In this appendix, we will give a formula 
associated with the restriction from $B_N(m,n)$ to $B_N(m,n-1)$. 
We will embed the subalgebra into $B_N(m,n)$ by removing the last 
slot of the $(m+n)$ slots. 
Under the embedding 
the elements of $B_N(m,n)$ can be expressed 
in terms of the elements of the subalgebra $B_N(m,n-1)$ as 
\begin{eqnarray}
b_i=\{
c_a, (\bar{j}\bar{n})c_a , c_a C_{k,n}
\}
\label{B(m,n)toB(m,n-1)},
\end{eqnarray}
where the index $i$ runs over a basis of $B_N(m,n)$, 
$i=1,\cdots,(m+n)!$, and 
$j=1,\cdots,n-1$ and $k=1,\cdots,m$. 
We have denoted the elements of $B_N(m,n-1)$ by $c_a$, 
where $a=1,\cdots,(m+n-1)!$.

We denote the dual element in $B_N(m,n)$ by $b^{\ast}$ 
and the dual element in $B_N(m,n-1)$ by $c^{\tilde{\ast}}$, 
which are defined by  
\begin{eqnarray}
\Sigma( b_i^{\ast})=\frac{1}{N^{m+n}}\Omega_{m+n}^{-1}
(\Sigma(b_i))^{-1}, 
\label{def:b_iast}
\end{eqnarray}
and 
\begin{eqnarray}
\Sigma( c_a^{\tilde{\ast}})=\frac{1}{N^{m+n-1}}\Omega_{m+n-1}^{-1}
(\Sigma(c_a))^{-1}.
\label{def:c_aast}
\end{eqnarray}
Note that $\Sigma$ is a map from $B_{N}(m,n)$ 
to $S_{m+n}$.

Using (\ref{def:b_iast}) for $b_i=c_a$, 
we get 
\begin{eqnarray}
\Sigma( c_a^{\ast})
&=&
\frac{1}{N^{m+n}}\Omega_{m+n}^{-1}
(\Sigma(c_a))^{-1}
\nonumber \\
&=&
\frac{1}{N}\Omega_{m+n}^{-1}\Omega_{m+n-1}
\Sigma (c_a^{\tilde{\ast}}).
\label{reduction1}
\end{eqnarray}
In the last step we have used (\ref{def:c_aast}). 
Note that $\Omega_{m+n-1}$ is the Omega factor 
defined in $S_{m+n-1}$ embedded in $S_{m+n}$ by removing 
the $(m+n)$-th slot.  

Similarly, for $b_i=(\bar{j}\bar{n})c_a$, we have 
\begin{eqnarray}
\Sigma( ( (\bar{j}\bar{n})c_a )^{\ast})
&=&
\frac{1}{N^{m+n}}\Omega_{m+n}^{-1}
\left(\Sigma((\bar{j}\bar{n})c_a)\right)^{-1}
\nonumber \\
&=&
\frac{1}{N^{m+n}}\Omega_{m+n}^{-1}
\left(\Sigma(c_a)(jn)\right)^{-1}
\nonumber \\
&=&
\frac{1}{N^{m+n}}\Omega_{m+n}^{-1}(jn)
\left(\Sigma(c_a)\right)^{-1}
\nonumber \\
&=&
\frac{1}{N}
(jn)
\Omega_{m+n}^{-1}\Omega_{m+n-1}
\Sigma (c_a^{\tilde{\ast}}), 
\label{reduction2}
\end{eqnarray}
and for $b_i=c_a C_{kn} $ we have 
\begin{eqnarray}
\Sigma( ( c_a C_{kn} )^{\ast})
&=&
\frac{1}{N^{m+n}}\Omega_{m+n}^{-1}
(\Sigma(c_a C_{kn}))^{-1}
\nonumber \\
&=&
\frac{1}{N^{m+n}}\Omega_{m+n}^{-1}
(\Sigma(c_a)( kn))^{-1}
\nonumber \\
&=&
\frac{1}{N^{m+n}}\Omega_{m+n}^{-1}(kn)
(\Sigma(c_a))^{-1}
\nonumber \\
&=&
\frac{1}{N}
(k n)
\Omega_{m+n}^{-1}\Omega_{m+n-1}
\Sigma (c_a^{\tilde{\ast}}).
\label{reduction3}
\end{eqnarray}
Note that we have used the following properties that come from 
the definition of $\Sigma$,
\begin{eqnarray}
&&
\Sigma(c_a C_{kn})=\Sigma(c_a ) (kn), 
\nonumber \\
&&
\Sigma((\bar{j}\bar{n}) c_a )=\Sigma(c_a ) (jn). 
\end{eqnarray}

Now
combining 
(\ref{reduction1}), 
(\ref{reduction2}) and (\ref{reduction3}), 
for $O_{c_a}^{\ast}:=Nc_a^{\ast}+
\sum_{j=1}^{n-1}\left((\bar{j}\bar{n})c_a\right)^{\ast}
+\sum_{k=1}^{m}\left(c_a C_{kn}\right)^{\ast}$ we have an interesting equation:
\footnote{
Note the following
\begin{eqnarray}
\left(
1+\frac{1}{N}\sum_{i\neq n}(in)
\right)
\Omega_{m+n-1}
=
\Omega_{m+n},
\end{eqnarray}
which appeared in \cite{9412110}. It was also exploited in 
\cite{0709.2158}. 
} 
\begin{eqnarray}
\Sigma( O_{c_a} ^{\ast})
&=&
\left(
1+\frac{1}{N}\sum_{i\neq n}(in)
\right)
\Omega_{m+n}^{-1}\Omega_{m+n-1}
(\Sigma (c_a^{\tilde{\ast}}))
\nonumber \\
&=&
\Sigma (c_a^{\tilde{\ast}}), 
\end{eqnarray}
which also means 
\begin{eqnarray}
 O_{c_a}^{\ast}
=
c_a^{\tilde{\ast}}. 
\label{the_formula}
\end{eqnarray}
Note that $c_a^{\ast}$ is not an element in $B_N(m,n-1)$, 
but $c_a^{\tilde{\ast}}$ is an element in $B_N(m,n-1)$.

For the reduction $B_N(2,1)\rightarrow B_N(1,1)$, 
we can confirm the following equation 
by computing the dual element explicitly:
\begin{eqnarray}
N1^{\ast}+(12)^{\ast}+C_{21}^{\ast} 
 =
\frac{1}{N^2-1}
\left(1-\frac{C_{11}}{N}\right)=1^{\tilde{\ast}},
\end{eqnarray}
where $1^{\tilde{\ast}}$ is defined in $B_N(1,1)$.
We can also find explicitly that 
the formula is valid in 
the reduction $B_N(2,1)\rightarrow S_2$ :
\begin{eqnarray}
 N1^{\ast}+C_{11}^{\ast}+C_{21}^{\ast}
 =
\frac{1}{N^2-1}
\left(1-\frac{s}{N}\right)
=\frac{1}{N^2+Ns} 
=\frac{1}{N^2\Omega_2} =1^{\tilde{\ast}},
\end{eqnarray}
where $1^{\tilde{\ast}}$ is defined in $S_2$. 


\section{Derivation of (\ref{first_terms_D4})}
\label{two-loop_detail}
\setcounter{equation}{0}

In this appendix we present explicit calculation to derive
(\ref{first_terms_D4}). 
The first term in 
(\ref{two-loop-dilatation}) is 
\begin{eqnarray}
&&tr(:[[Y,X],\partial_X][[\partial_Y,\partial_X],X]:)
\nonumber \\
&=&
tr(:[Y,X]\partial_X [\partial_Y,\partial_X]X:)
-tr(:[Y,X]\partial_X X[\partial_Y,\partial_X]:)
\nonumber \\
&&
-tr(:\partial_X [Y,X][\partial_Y,\partial_X]X:)
+tr(:\partial_X [Y,X]X[\partial_Y,\partial_X]:).
\label{twoloopdilatation-1}
\end{eqnarray}
In what follows we will not write the normal ordering symbol explicitly. 
Using $(\partial_X)_j^i X^k_l=\delta^i_l\delta_j^k $, 
\begin{eqnarray}
&&
tr(A\partial_X\partial_Y\partial_X) O^{\gamma}_{A,ij}(X,Y)
\nonumber \\
&=&
(A)^a_b (\partial_X)^b_c(\partial_Y)^c_d(\partial_X)^d_a 
(Q^{\gamma}_{A,ij})^{JL}_{IK}X^I_J (Y^T)^K_L
\nonumber \\
&=&
m(m-1)n
(A)^a_b 
(Q^{\gamma}_{A,ij})^{d b j_3 \cdots d l_2 \cdots}_{ac i_3 \cdots c k_2 \cdots}
X^{i_3 \cdots }_{j_3 \cdots} (Y^T)^{k_2 \cdots}_{l_2 \cdots}
\nonumber \\
&=&
m(m-1)n
tr_{m,n}(Q^{\gamma}_{A,ij} (12)C_{1,1} 1\otimes A\otimes X^{\otimes m-2}\otimes 1\otimes 
Y^{T}{}^{\otimes n-1})
\nonumber \\
&=&
m(m-1)n
tr_{m,n}( C_{1,1}(12)Q^{\gamma}_{A,ij} 1 \otimes A
\otimes X^{\otimes m-2}\otimes 1\otimes 
Y^{T}{}^{\otimes n-1}),
\end{eqnarray}
where $A=X[Y,X]$. 
Likewise, we have
\begin{eqnarray}
&&
tr(A\partial_X\partial_X\partial_Y) O^{\gamma}_{A,ij}(X,Y)
\nonumber \\
&=&
m(m-1)n
(A)^a_b 
(Q^{\gamma}_{A,ij})^{c b j_3 \cdots a l_2 \cdots}_{dc i_3 \cdots d k_2 \cdots}
X^{i_3 \cdots }_{j_3 \cdots} (Y^T)^{k_2 \cdots}_{l_2 \cdots}
\nonumber \\
&=&
m(m-1)n
tr_{m,n}(Q^{\gamma}_{A,ij} C_{1,1}(12) 
1\otimes A\otimes X^{\otimes m-2}\otimes 1\otimes 
Y^{T}{}^{\otimes n-1}).
\end{eqnarray}
Then 
one finds that the first line in 
(\ref{twoloopdilatation-1}) can be expressed by 
\begin{eqnarray}
m(m-1)n
tr_{m,n}([C_{1,1}(12),Q^{\gamma}_{A,ij}]  
1\otimes X[Y,X]\otimes X^{\otimes m-2}\otimes 1\otimes 
Y^{T}{}^{\otimes n-1}).
\end{eqnarray}

We next calculate 
\begin{eqnarray}
&&
tr(B\partial_X X\partial_Y\partial_X) O^{\gamma}_{A,ij}(X,Y)
\nonumber \\
&=&
(B)^a_b (\partial_X)^b_c (X)^c_d (\partial_Y)^d_e(\partial_X)^e_a 
(Q^{\gamma}_{A,ij})^{JL}_{IK}X^I_J (Y^T)^K_L
\nonumber \\
&=&
m(m-1)n
(B)^a_b (X)^c_d
(Q^{\gamma}_{A,ij})^{e b j_3 \cdots e l_2 \cdots}_{ac i_3 \cdots d k_2 \cdots}
X^{i_3 \cdots }_{j_3 \cdots} (Y^T)^{k_2 \cdots}_{l_2 \cdots}
\nonumber \\
&=&
m(m-1)n
tr_{m,n}((12)C_{1,1}Q^{\gamma}_{A,ij} B\otimes  X^{\otimes m-1}\otimes 1\otimes 
Y^{T}{}^{\otimes n-1} ), 
\end{eqnarray}
where $B=[Y,X]$, and 
\begin{eqnarray}
&&
tr(B\partial_X X\partial_X\partial_Y) O^{\gamma}_{A,ij}(X,Y)
\nonumber \\
&=&
m(m-1)n
(B)^a_b (X)^c_d
(Q^{\gamma}_{A,ij})^{d b j_3 \cdots a l_2 \cdots}_{ec i_3 \cdots e k_2 \cdots}
X^{i_3 \cdots }_{j_3 \cdots} (Y^T)^{k_2 \cdots}_{l_2 \cdots}
\nonumber \\
&=&
m(m-1)n
tr_{m,n}(Q^{\gamma}_{A,ij} (12)C_{1,1} B\otimes  X^{\otimes m-1}\otimes 1\otimes 
Y^{T}{}^{\otimes n-1} ). 
\label{re-writing1}
\end{eqnarray}
From these two,
one finds that the second term in 
(\ref{twoloopdilatation-1}) can be expressed by 
\begin{eqnarray}
-m(m-1)n
tr_{m,n}([(12)C_{1,1},Q^{\gamma}_{A,ij}]  B\otimes  X^{\otimes m-1}\otimes 1\otimes 
Y^{T}{}^{\otimes n-1} ).
\end{eqnarray}

Similarly, we have 
\begin{eqnarray}
&&
tr(B\partial_Y \partial_X X\partial_X) O^{\gamma}_{A,ij}(X,Y)
\nonumber \\
&=&
m(m-1)n
(B)^a_b (X)^d_e
(Q^{\gamma}_{A,ij})^{e c j_3 \cdots c l_2 \cdots}_{ad i_3 \cdots b k_2 \cdots}
X^{i_3 \cdots }_{j_3 \cdots} (Y^T)^{k_2 \cdots}_{l_2 \cdots}
\nonumber \\
&=&
m(m-1)n
tr_{m,n}(C_{1,1}(12) Q^{\gamma}_{A,ij}  B\otimes  X^{\otimes m-1}\otimes 1\otimes 
Y^{T}{}^{\otimes n-1}), 
\label{re-writing2}
\end{eqnarray}
and 
\begin{eqnarray}
&&
tr(B\partial_X \partial_Y X\partial_X) O^{\gamma}_{A,ij}(X,Y)
\nonumber \\
&=&
m(m-1)n
(B)^a_b (X)^d_e
(Q^{\gamma}_{A,ij})^{e b j_3 \cdots d l_2 \cdots}_{ac i_3 \cdots c k_2 \cdots}
X^{i_3 \cdots }_{j_3 \cdots} (Y^T)^{k_2 \cdots}_{l_2 \cdots}
\nonumber \\
&=&
m(m-1)n
tr_{m,n}(Q^{\gamma}_{A,ij}C_{1,1}(12)  B\otimes  X^{\otimes m-1}\otimes 1\otimes 
Y^{T}{}^{\otimes n-1}). 
\end{eqnarray}
One then finds that the third term in 
(\ref{twoloopdilatation-1}) can be expressed by 
\begin{eqnarray}
-m(m-1)n
tr_{m,n}([C_{1,1}(12),Q^{\gamma}_{A,ij}]  B\otimes  X^{\otimes m-1}\otimes 1\otimes 
Y^{T}{}^{\otimes n-1} ). 
\end{eqnarray}

Finally we have 
\begin{eqnarray}
&&
tr(\tilde{A}\partial_Y\partial_X\partial_X) O^{\gamma}_{A,ij}(X,Y)
\nonumber \\
&=&
m(m-1)n
(\tilde{A})^a_b 
(Q^{\gamma}_{A,ij})^{cd j_3 \cdots c l_2 \cdots}_{da i_3 \cdots b k_2 \cdots}
X^{i_3 \cdots }_{j_3 \cdots} (Y^T)^{k_2 \cdots}_{l_2 \cdots}
\nonumber \\
&=&
m(m-1)n
tr_{m,n}((12)C_{1,1}Q^{\gamma}_{A,ij} 
1\otimes \tilde{A}\otimes X^{\otimes m-2}\otimes 1\otimes 
Y^{T}{}^{\otimes n-1}  ), 
\end{eqnarray}
where $\tilde{A}=[Y,X]X$, and
\begin{eqnarray}
&&
tr(\tilde{A}\partial_X\partial_Y\partial_X) O^{\gamma}_{A,ij}(X,Y)
\nonumber \\
&=&
m(m-1)n
tr_{m,n}(Q^{\gamma}_{A,ij} (12) C_{11}
1\otimes \tilde{A}\otimes X^{\otimes m-2}\otimes 1\otimes 
Y^{T}{}^{\otimes n-1}  )
\end{eqnarray}
to give the following expression for the last term in 
(\ref{twoloopdilatation-1}):
\begin{eqnarray}
m(m-1)n
tr_{m,n}([(12)C_{1,1},Q^{\gamma}_{A,ij}] 
1\otimes \tilde{A}\otimes X^{\otimes m-2}\otimes 1\otimes 
Y^{T}{}^{\otimes n-1}  ). 
\end{eqnarray}


\end{document}